\documentstyle[preprint,aps]{revtex}

\begin{document}


\def \XX{{\widehat{\cal I}}}
\def \GeV{{\rm \enspace GeV}}
\def \TeV{{\rm \enspace TeV}}
\def \beq{\begin{equation}}
\def \eeq{\end{equation}}
\def \beqa{\begin{eqnarray}}
\def \eeqa{\end{eqnarray}}
\def \half{\hbox{$1\over2$}}
\def \mt{m_t}
\def \ceb{ {\hat{c}}_{\bar{e}b} }
\def \cmub{ {\hat{c}}_{\mu\bar{b}} }
\def \sqt{ s_{qt} }
\def \cqt{ c_{qt} }
\def \cet{ c_{\bar{e}t} }
\def \set{ s_{\bar{e}t} }
\def \cmut{ c_{\mu\bar{t}} }
\def \smut{ s_{\mu\bar{t}} }
\def \ceq{ c_{\bar{e}q} }
\def \seq{ s_{\bar{e}q} }
\def \cmuq{ c_{\mu\bar{q}} }
\def \smuq{ c_{\mu\bar{q}} }
\def \cemu{ c_{\bar{e}\mu} }
\def \ME{ {\cal{M}} }
\def \PROD{ \sum{\cal M}_{\lambda\bar\lambda}
            {\cal M}^{*}_{\lambda^\prime\bar\lambda^\prime} }


\draft
\preprint{
  \parbox{2in}{Fermilab--Pub--97/185-T \\
  [-0.12in] UM--TH--97--09 \\
  [-0.12in] hep-ph/9706304
}  }

\title{Maximizing Spin Correlations in\\
Top Quark Pair Production at the Tevatron}
\author{Gregory Mahlon  \cite{GDMemail}}
\address{Department of Physics, University of Michigan \\
500 E. University Ave., Ann Arbor, MI  48109 }
\author{Stephen Parke \cite{SPemail}}
\address{Fermi National Accelerator Laboratory \\
P.O. Box 500, Batavia, IL  60510 }
\date{June 6, 1997}
\maketitle
\begin{abstract}
A comparison is made between the off-diagonal and helicity
spin bases for top quark pair production at the FNAL Tevatron.
In the off-diagonal basis, 92\% of the top quark pairs are 
in the spin configuration up-down plus down-up, whereas in the
helicity basis only 70\% are left-right plus right-left.
The off-diagonal basis maximizes the spin asymmetry and
hence the measured angular correlations between the decay products, 
which are more than twice as big in this basis as compared to the 
helicity basis.  In addition, for the process
$q\bar{q}\rightarrow t\bar{t}$,
we give a very simple analytic expression for the matrix element 
squared which includes {\it all}\ spin correlations between the 
production and subsequent decay of the top quarks.
\end{abstract}


Since the discovery~\cite{TopExpt} of the top quark in 1995,
a number of authors~\cite{MP,SW,other} have revisited the question
of spin correlations in top quark pair production at the FNAL Tevatron.
(References to the extensive literature of
earlier works can be found in these recent papers.)
With its high mass of about 175 GeV,
the top quark decays before it hadronizes~\cite{Bigi}.  Thus,
the decay products of a top quark produced in a definite spin state
will have characteristic angular correlations.
The reason for these new studies is the realization that the
number of like-spin and unlike-spin top quark pairs can be made
significantly different by an appropriate choice of spin basis.
Both Mahlon and Parke~\cite{MP} and Stelzer and Willenbrock~\cite{SW}
discussed the spin asymmetry using the helicity basis for the Tevatron.
However, the top quarks produced at the Tevatron are not
ultra-relativistic; therefore, the helicity basis is probably not the
optimal basis for these studies at this machine.
Consequently, Ref.~\cite{MP} went on to consider
the beamline basis, which is more
suitable for spin studies near threshold, and found larger effects
at the Tevatron in this basis.
The question remained, however: what is the optimal spin basis for
these correlation studies?  
We now have an answer to this question.

Recently Parke and Shadmi~\cite{PS}
studied the spin correlations for the process
$e^+e^- \rightarrow t \bar{t}$ and showed that
you can choose a spin basis
in which the like spin components, up-up (UU) and down-down (DD), 
identically vanish. 
They have called this spin basis the off-diagonal basis.
In a footnote these authors briefly applied this result to 
$q \bar{q} \rightarrow t \bar{t}$, 
which is the dominant top quark production process at the Tevatron. 
The purpose of this letter is to study the 
$q \bar{q} \rightarrow t \bar{t}$
process in more detail, comparing the off-diagonal basis to the more
traditional helicity basis. 
Although the type of analysis described in Ref.~\cite{PS} can
also be applied to the $gg$ initial state, the
dominant $t\bar{t}$ production mechanism at the LHC, 
we find that no significant improvements 
over the helicity basis are possible at the LHC.
At the Tevatron\footnote{Throughout this paper, we take the center
of mass energy for the $p\bar{p}$ collisions to be 2.0 TeV.
We use a top quark mass of 175 GeV, $W$ boson mass 
of 80 GeV, and employ the 
MRS(R1) structure functions~\protect\cite{NewPDF} evaluated
at the scale $Q^2 = m_W^2$.}
however, the process
$q\bar{q}\rightarrow t \bar{t}$ dominates,
accounting for approximately
88\%  of the total $t\bar{t}$ cross section.
Thus, we will concentrate our analytic discussions on
that process.  Our numerical results, however, include
both the $q\bar{q}$ and $gg$ initial states.

In this letter, we present the production density matrix for 
$q \bar{q} \rightarrow t\bar{t}$ in the off-diagonal basis 
and contrast it with the known result for the helicity basis.
Then we obtain a surprisingly simple and compact
expression for the production and decay of a pair of top quarks
from a quark-antiquark initial state which 
includes {\it all}\ of the correlations among the particles.
We also show that the interference terms for the off-diagonal basis
are substantially smaller than for the helicity basis at the Tevatron.
Finally, we apply this basis to spin correlation studies at the 
Tevatron, where we find that in the off-diagonal basis, 
92\% of the $t\bar{t}$ pairs produced at the Tevatron have unlike 
spins.  This represents a significant improvement over 
the helicity basis, in which only 70\% of the $t\bar{t}$ pairs 
have unlike helicities.  As a result, the correlations obtained in
the off-diagonal basis are more than twice as big as those
in the helicity basis.

\begin{figure}[h]

\vspace*{15cm}
\includegraphics{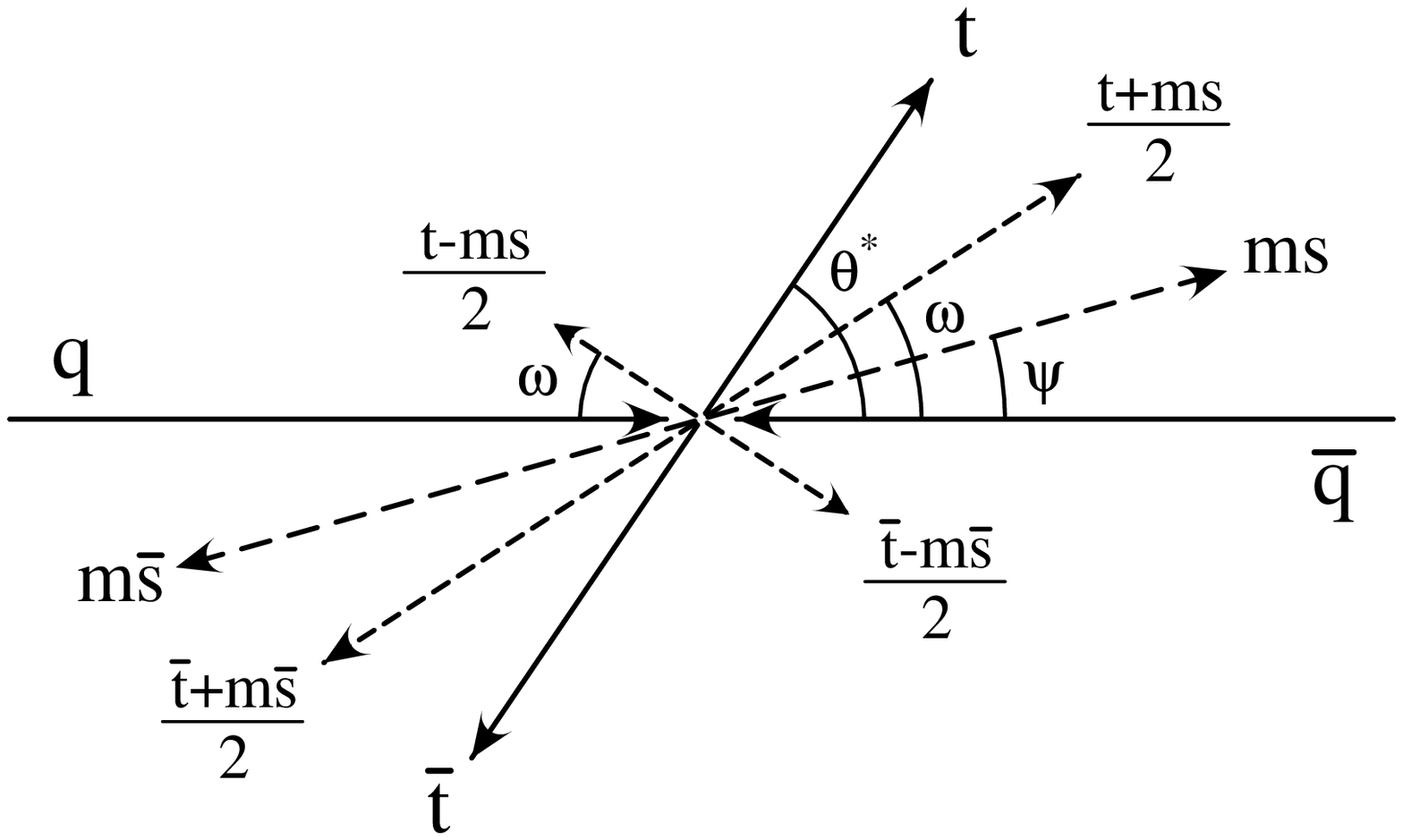}
\vspace{1.0cm}

\caption[]{The relevant angles and vectors 
in the zero momentum frame of the initial $q\bar{q}$ pair
for the off-diagonal basis of Parke and Shadmi.  
The top quark is produced at an angle $\theta^*$ with
respect to the beam axis ($\cos\theta^* \equiv c_{qt}$).
The spin
vector $s$ makes an angle $\psi$ (given by Eq.~(\protect\ref{psidefn}))
with respect to the beam axis.  The vectors $(t\pm ms)/2$,
where $m$ is the top quark mass, indicate
the preferred emission directions for the charged lepton or down-type
quark from the decaying $W^{+}$ (see Eq.~(\protect\ref{omegadefn})).  
The vectors describing the antitop
lie back-to-back with the corresponding top quark vectors.
}
\label{ANGLES}
\end{figure}

In Fig.~\ref{ANGLES} the spatial part of the 
top quark momenta ($t, \bar{t}\thinspace$) 
and spin vectors ($s, \bar{s}$) are given for the process 
$q \bar{q} \rightarrow t \bar{t}$
in the zero momentum frame (ZMF) of the incoming quarks.
What was shown in Ref.~\cite{PS} is that the spin projections 
of the top-antitop pair in $q \bar{q} \rightarrow t \bar{t}$
are purely up-down and down-up if the spin vectors 
make an angle $\psi$ with respect to the beam axis.  
This angle is given by
\beq
\tan \psi = {{ \beta^2 \sqt \cqt} \over	{1 - \beta^2 \sqt^2}},
\label{psidefn}
\eeq
where $\beta$ is the speed of the top quarks in 
the ZMF.\footnote{The angle $\psi$ is related to the angle
$\alpha$ introduced by Tsai~\protect\cite{Tsai} for
$e^{+}e^{-}\rightarrow \tau^{+}\tau^{-}$.}
Throughout this paper
we use the notation $s_{ij}$ and $c_{ij}$ to denote the sine
and cosine of the angle between the momenta of
particles $i$ and $j$ in the ZMF:  hence, $c_{qt}$ is the cosine of the 
top quark scattering angle (often denoted by $\cos \theta^{*}$).
Note that near threshold the spin vectors are aligned along 
the beam direction and that at very high energies they are aligned 
along the direction of the top and antitop momenta.

For the off-diagonal basis, the production density matrix
averaged over the color and spin of the incoming quarks
is given by\footnote{
The phases of the non-diagonal terms of the production density matrix
are dependent on many of our conventions, which we have chosen
to make the phases as simple as possible.
We take the 31 plane to be the scattering
plane with the top quark direction the 1 axis. 
We use the conventions of \protect\cite{MP} with the spinor products
singular along the minus 2 direction and 
$ \langle k + | q - \rangle \equiv
\Bigl\{
\bigl[ q_3(k_0{+}k_2) - k_3(q_0{+}q_2) \bigr] 
+i\bigl[ k_1(q_0{+}q_2) - q_1(k_0{+}k_2) \bigr] 
\Bigr\}\Bigl\{(k_0{+}k_2)(q_0{+}q_2)\Bigr\}^{-1/2}  $
for positive energy light-like vectors $k$ and $q$.
}
\beq
\PROD = { g_s^4 \over 9}
\left[
\begin{array}{cccc}
0\enspace & 0                 & 0                 &\enspace 0 \\
0\enspace & 2{-}\beta^2\sqt^2 & \beta^2\sqt^2     &\enspace 0 \\
0\enspace & \beta^2\sqt^2     & 2{-}\beta^2\sqt^2 &\enspace 0 \\
0\enspace & 0                 & 0                 &\enspace 0 
\end{array}
\right].
\label{Off}
\eeq
The matrix element for
the production of a $t$ quark with spin $\lambda$ and a $\bar{t}$
quark with spin $\bar\lambda$ is ${\cal M}_{\lambda\bar\lambda}$, and
$g_s$ is the strong coupling constant.
The ordering of the columns and rows in~(\ref{Off})
is (UU,UD,DU,DD).  This production density matrix is 
very simple because in the off-diagonal basis the
amplitudes with like spins, UU and DD, vanish identically.
Also note that 
the non-diagonal terms have an explicit factor of $\beta^2$.

In contrast, the same production density matrix in terms of
the helicity basis reads~\cite{Lee}
\beq
\PROD = {g_s^4 \over 9 }
\left[
\begin{array}{cccc}
\sqt^2 / \gamma^2 \enspace  & -\sqt\cqt / \gamma \enspace & 
\sqt\cqt / \gamma \enspace  & \sqt^2 / \gamma^2           \\
-\sqt\cqt / \gamma\enspace  & 2-\sqt^2 \enspace           & 
\sqt^2            \enspace  & -\sqt\cqt / \gamma          \\
\sqt\cqt / \gamma \enspace  & \sqt^2  \enspace            & 
2-\sqt^2          \enspace  & \sqt\cqt / \gamma           \\
\sqt^2 / \gamma^2 \enspace  & -\sqt\cqt / \gamma \enspace & 
\sqt\cqt / \gamma \enspace  & \sqt^2 / \gamma^2
\end{array}
\right],
\label{Hel}
\eeq
where $\gamma = (1{-}\beta^2)^{-1/2}$ is the usual Lorentz boost
factor.  The columns and rows of this matrix have been ordered 
(RR,RL,LR,LL), with obvious notations for right and left helicities.

In the off-diagonal spin basis, the non-diagonal terms of the production
density matrix are at least a factor of $\beta^2$ times smaller than
for the helicity basis.  For the Tevatron at 2~TeV this corresponds 
to a factor of typically 0.3 to 0.4.
Furthermore, the terms lying on the edges of the production
density matrix in the helicity basis are not very strongly
suppressed compared to the diagonal terms, 
as $\gamma$ is only 1.2 to 1.3 at these values of $\beta^2$.
These properties of the two matrices
translate into smaller interference terms in the off-diagonal basis
once the decays are included.

Using either of the above production density matrices and 
the corresponding decay density matrices, the total
matrix element squared
for the production and decay process\footnote{
We have assumed that the decays of both $W$-bosons are leptonic.
To change one or both of the $W$-boson decays to a hadronic decay,
replace the charged lepton with a down type quark and the neutrino
with an up type quark. This will preserve all correlations.}
$q \bar{q} \rightarrow t \bar{t} 
\rightarrow W^+ b ~ W^- \bar{b} 
\rightarrow \bar{e} \nu b ~\mu \bar{\nu} \bar{b}$
averaged over the initial quark's color and spin
and summed over the final colors and spins
is given by
\beqa
\sum_{}^{} &&
\vert \ME \vert^2 = 
\nonumber\\ &&
{g_s^4 \over 9} {\cal T}{\bar{\cal T}}
\Biggl\{
(2-\beta^2 \sqt^2)
- {
{ (1{-}\ceq\cmuq) 
  - \beta(\cmut{+}\cet)
  + \beta\cqt(\ceq{+}\cmuq) 
  + \half\beta^2 \sqt^2 (1{-}\cemu) }
\over
{ \gamma^2(1-\beta\cet)(1-\beta\cmut) }
}
\Biggr\}.
\label{Fullmsq}
\eeqa
The factor $\cal T$ comes from the decay of the top quark
($t \rightarrow W^+ b \rightarrow \bar{e} \nu_e b$):
\beq
{\cal T} =
{ {g_W^4} \over {4m_t^2 \Gamma_t^2} }
\thinspace
( m_t^2-2\bar{e}\cdot\nu )
\thinspace
{
{\mt^2(1-\ceb^2) + (2\bar{e}\cdot\nu)(1+\ceb)^2}
\over
{ (2\bar{e}\cdot\nu-m_W^2)^2+(m_W\Gamma_W)^2 }
},
\label{Tdef}
\eeq
where $\ceb$ is the cosine of angle between $\bar{e}$ and $b$ in 
the $W^{+} (=\bar{e} + \nu)$ rest frame, 
$2\bar{e}\cdot\nu$ is the invariant 
mass of the positron and neutrino,
$(m_t, \Gamma_t)$ and $(m_W, \Gamma_W)$ are the masses and widths
of the top quark and $W$-boson respectively, 
and $g_W$ is the weak coupling constant.
Apart from the factor $(m_t\Gamma_t)^{-2}$, which comes from
the top quark propagator, ${\cal T}$ is just the 
matrix element squared for unpolarized top quark decay.
Likewise, $\bar{\cal T}$ is from the antitop decay 
($\bar{t} \rightarrow W^- \bar{b} \rightarrow 
\mu {\bar{\nu}}_\mu \bar{b}$):
\beq
\bar{\cal T} =
{ {g_W^4} \over {4m_t^2 \Gamma_t^2} }
\thinspace
( m_t^2-2\mu\cdot\bar\nu )
\thinspace
{
{\mt^2(1-\cmub^2) + (2\mu\cdot\bar\nu)(1+\cmub)^2}
\over
{ (2\mu\cdot\bar\nu-m_W^2)^2+(m_W\Gamma_W)^2 }
},
\label{Tbardef}
\eeq
where
$\cmub$ is the cosine of angle between $\mu$ and $\bar{b}$ in 
the $W^{-} (=\mu + \bar{\nu})$ rest frame,
and $2\mu\cdot\bar{\nu}$ is the invariant 
mass of the muon and anti-neutrino.
Eq.~(\ref{Fullmsq}) agrees with the results 
of Kleiss and Stirling\cite{ks}
for on mass shell top quarks and  massless $b$-quarks\footnote{
Inclusion of the finite mass effects for the
$b$-quarks would result in straightforward but messy modifications to
Eqs.~(\protect\ref{Tdef}) and~(\protect\ref{Tbardef}).
The size of these effects is typically less than 1\%.}
independent of whether or not the $W$-bosons are on or off mass shell.

If we did not include the  
spin correlations between production and decay,
the total matrix element squared would be simply
$g_s^4 {\cal T}{\bar{\cal T}} (2-\beta^2 \sqt^2)/9$.
Thus, all of the correlations between the production and decay
of the top quarks are contained in the second term inside the braces
of Eq.~(\ref{Fullmsq}).  It is noteworthy that of the six final
state particles, only the directions of the two charged leptons
are required in addition to the directions of the $t$ and $\bar{t}$
to fully specify these correlations.  
For the up-down spin configuration, the preferred emission 
directions for the charged leptons are $(t+ms)/2$ for 
the positron and $(\bar{t}+m\bar{s})/2$ for the muon:  
for the down-up configuration they are $(t-ms)/2$ and 
$(\bar{t}-m\bar{s})/2$ (see Fig.~\ref{ANGLES}).
These light-like vectors make an angle $\omega$ with respect 
to the beam axis given by
\beq
\sin \omega = \beta \sqt,
\label{omegadefn}
\eeq
and their energy components are 
$\gamma m_t(1\pm\beta\cqt\sec\omega)/2$.

\begin{figure}[h]

\vspace*{15cm}
\includegraphics{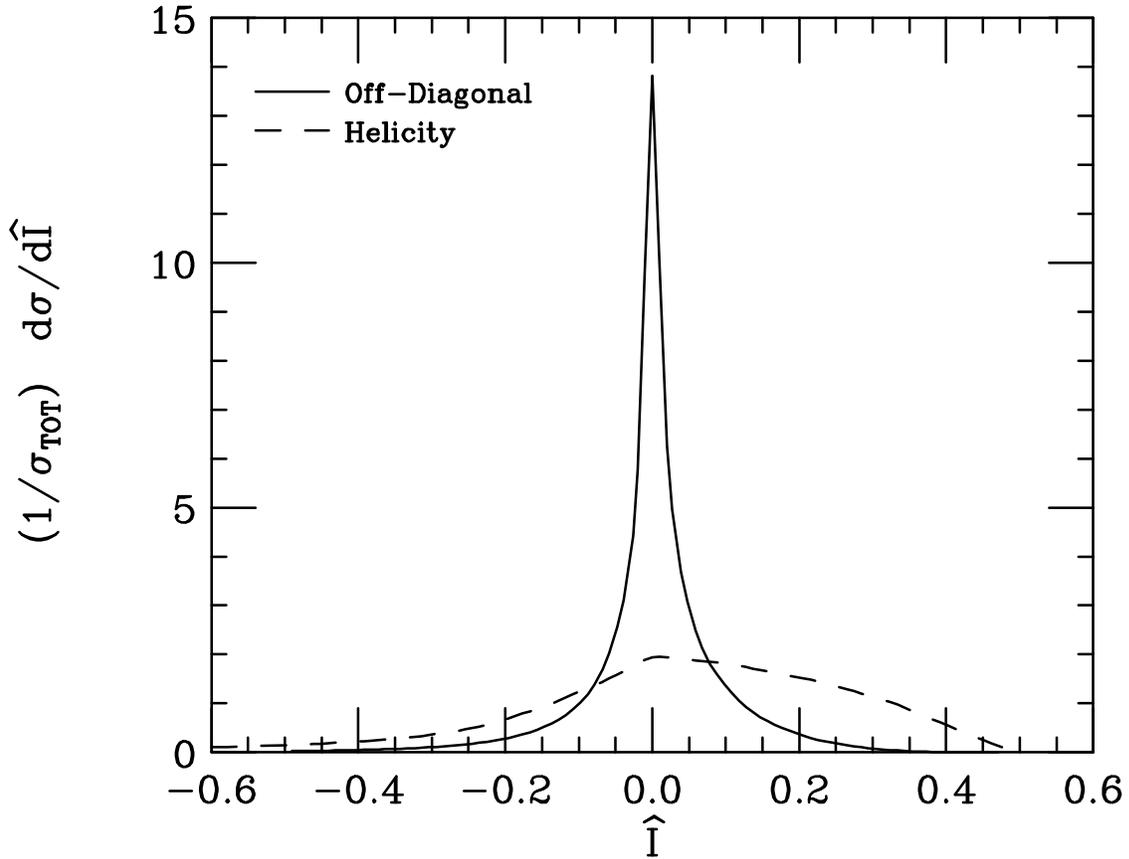}
\vspace{1.0cm}

\caption[]{The relative importance of the interference terms
in the off-diagonal and helicity  bases in $q\bar{q}\rightarrow t\bar{t}$
for the Tevatron at $\sqrt{s} = 2 \TeV$.
Plotted is the differential distribution in 
$\XX\equiv {\cal I}/\sum\vert{\cal M}\vert^2$, 
the value of the interference
term (Eq.~(\protect\ref{Iusb}) or~(\protect\ref{Ihel})) normalized
to the square of the total matrix element 
(Eq.~(\protect\ref{Fullmsq})).  In the off-diagonal
basis, 90\% of the cross section comes from phase space points
where $\vert \XX \vert < 0.15$,  
whereas in the helicity basis only 50\% of the cross section
comes from this region.
}
\label{INFPLOT}
\end{figure}

In the off-diagonal spin basis
the interference terms ({\it i.e.} those terms in Eq.~(\ref{Fullmsq}) 
coming from the non-diagonal
pieces of the production density matrix) are
\beqa
{{\cal I}_{o}} =  
{
{g_s^4{\cal T}{\bar{\cal T}} }
\over
{9 \gamma^2(1-\beta\cet)(1-\beta\cmut) } 
}
\thinspace
{ {\beta^2} \over {2} }
&& \Bigl[
(\cet-\cqt\ceq-\beta\sqt^2)(\cmut-\cqt\cmuq-\beta\sqt^2) 
/(1-\beta^2\sqt^2) 
\nonumber\\[0.2cm] && \enspace +
 (\cet-\cqt\ceq)(\cmut-\cqt\cmuq) + \sqt^2(\cemu+\ceq\cmuq) 
\Bigr],
\label{Iusb}
\eeqa
whereas for the helicity basis the interference terms are 
\beqa
{{\cal I}_{h}} = 
{ 
{g_s^4{\cal T}{\bar{\cal T}}}
\over 
{9 \gamma^2(1-\beta\cet)(1-\beta\cmut) } 
} &&
\Bigl[
 (\beta\cqt - \ceq)(\beta\cqt - \cmuq)
\cr && \enspace
  -\cqt^2(\beta-\cet)(\beta-\cmut)
  +\half\beta^2\sqt^2(\cemu+\cet\cmut) 
\Bigr].
\label{Ihel}
\eeqa
Here we see again that the interference terms are a factor of $\beta^2$ 
smaller for the off-diagonal spin basis than the helicity basis.
This point may be illustrated by plotting the distribution in
$\XX \equiv {\cal I}/\sum\vert{\cal M}\vert^2$:  
for each phase space point
we compute the value of the interference term and divide by
the total matrix element squared at that point.  Because the 
total matrix element squared
can range from 0 (maximal destructive interference)
to $2{\cal I}$ (maximal constructive interference), 
the variable $\XX$ must
lie in the interval $(-\infty,{1\over2}]$.
The resulting
differential distributions for both bases at the Tevatron
is shown in Fig.~\ref{INFPLOT}.
In the off-diagonal basis, $d\sigma /d\XX$ resembles a sharp spike:  
in fact, 90\% of the cross section
comes from points where $\vert \XX \vert < 0.15$.  In contrast,
only about half of the cross section comes from this region in
the helicity basis.  To enclose 90\% of the cross section 
in the helicity basis, 
we must expand the range to $\vert \XX \vert < 0.35$.
Clearly, the interference terms are much less important
in the off-diagonal basis than in the helicity basis.
Thus, the off-diagonal basis provides a far superior
description of the $q \bar{q} \rightarrow t \bar{t}$
process at this accelerator.

\begin{figure}[h]

\vspace*{15cm}
\includegraphics{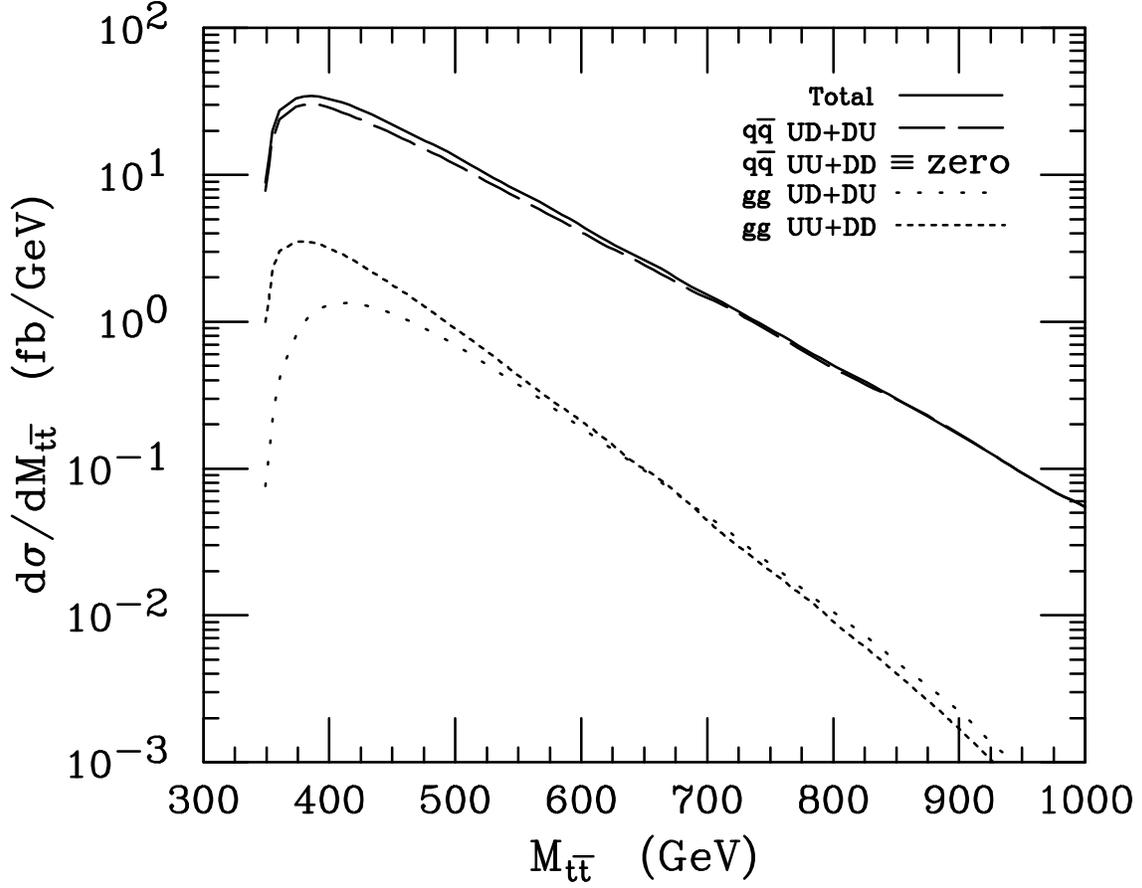}
\vspace{1.0cm}

\caption[]{Differential cross section for $t\bar{t}$ production
as a function of the $t\bar{t}$ invariant mass $M_{t\bar{t}}$
for the Tevatron with center-of-mass energy 2.0 TeV, decomposed
into UD+DU and UU+DD spins of the $t\bar{t}$ pair using
the off-diagonal basis for both $q\bar{q}$ and $gg$ components.
}
\label{MASSPLOT}
\end{figure}

In Fig.~\ref{MASSPLOT} we show the breakdown of the  total
$t\bar{t}$ cross section into like- and unlike-spin pairs
using the off-diagonal basis versus
the $t\bar{t}$ invariant mass for the
Tevatron.   We find that 92\% of the
$t\bar{t}$ pairs produced have unlike spins  (UD+DU) in
this basis.   In contrast, only 70\% of the $t\bar{t}$ pairs
have unlike helicities.\footnote{The value of 67\% unlike
helicities we quoted in Ref.~\protect\cite{MP} is based
on an older set of structure functions,
which contained a somewhat larger gluon component.
The distributions we consider are not significantly affected
by the choice of structure functions.}  Note that we have
chosen spin combinations which are insensitive to
which  beam donated the quark in $q\bar{q}\rightarrow t\bar{t}$.

To observe the resulting correlations between the decay 
products of the top and the antitop, we proceed as
described in Ref.~\cite{MP}.   Suppose that the $i$th
decay product of the top quark is emitted at an angle
$\theta_i$ with respect to the top spin axis 
in the top rest frame, and that the $\bar{\imath}$th decay
product of the antitop is emitted at an angle
$\bar\theta_{\bar{\imath}}$  with respect to the antitop
spin axis in the antitop rest frame.   We  tag the top quark
in a particular event as having spin up if 
$\alpha_i \cos\theta_i > 0$, and as having spin down otherwise.
The angular distribution of the $\bar{\imath}$th decay product
in this situation is
\beq
{1 \over \sigma_{\rm\scriptscriptstyle TOT}} ~
{ d \sigma \over d(\cos \bar\theta_{\bar{\imath}}) } = 
{1\over2}
\Bigl[ 1 + \hbox{$1\over2$}(1-2P_\times)\thinspace
           \alpha_i\alpha_{\bar{\imath}}\cos\bar\theta_{\bar{\imath}} 
\Bigr],
\label{CorrelEq}
\eeq
where 
$P_\times$ is the fractional purity of the unlike-spin component
of the sample of $t\bar{t}$ events.  The $\alpha_i$'s 
($\alpha_{\bar\imath}$'s) are the
correlation coefficients from the decay distribution of a polarized
top (antitop)~\cite{Spin} and take on the values
$\alpha_{\bar{e}} = \alpha_{\bar{d}} = 1$, 
$\alpha_{\nu} = \alpha_{u} = -0.31$,
and $\alpha_b = -0.41$ for the decay of a 175 GeV spin-up top quark.
The $\alpha_i$'s for spin-down top quarks have opposite sign.
The correlation coefficients for a spin up (spin down) antitop
are the same as for a spin down (spin up) top quark.

Because the $W$ boson has primarily hadronic decays, it is useful
to have a method to probabilistically determine which of the
two jets in such a decay was initiated by the down-type quark.
As explained in Ref.~\cite{MP}, the jet which lies closest
to the $b$-quark direction as viewed in the $W$ rest frame
is most likely the $d$-type quark.  The probability that this
identification is correct is
$P_d = {1\over4} ( 2m_t^2 + 7m_W^2 ) / (m_t^2 + 2m_W^2)$,
and equals 0.61.  
The effective correlation coefficient for this {\it ``d''}-type
quark is given by
$\alpha_{\it ``d"} = P_d \alpha_d + (1-P_d)\alpha_{\bar u}$,
or about $-0.49$.

Since the coefficient of $\cos\bar\theta_{\bar{\imath}}$ governs
the size of the observable correlations, it is desirable
to make it as large as possible.  Thus, we should choose
to work in the off-diagonal basis where $1-2P_\times$ is $-0.84$
instead of the helicity basis where $1-2P_\times$ is only $-0.39$.
The other two factors, 
$\alpha_i$ and $\alpha_{\bar{\imath}}$ depend on
which top decay product is used to tag the top quark spin,
and which antitop decay product angular distribution is generated.
The largest correlations are clearly between the two charged leptons.
For the same pair of decay products, the correlations
are more than  twice as large in the off-diagonal basis as in the
helicity basis.

\begin{figure}[h]

\vspace*{15cm}
\includegraphics{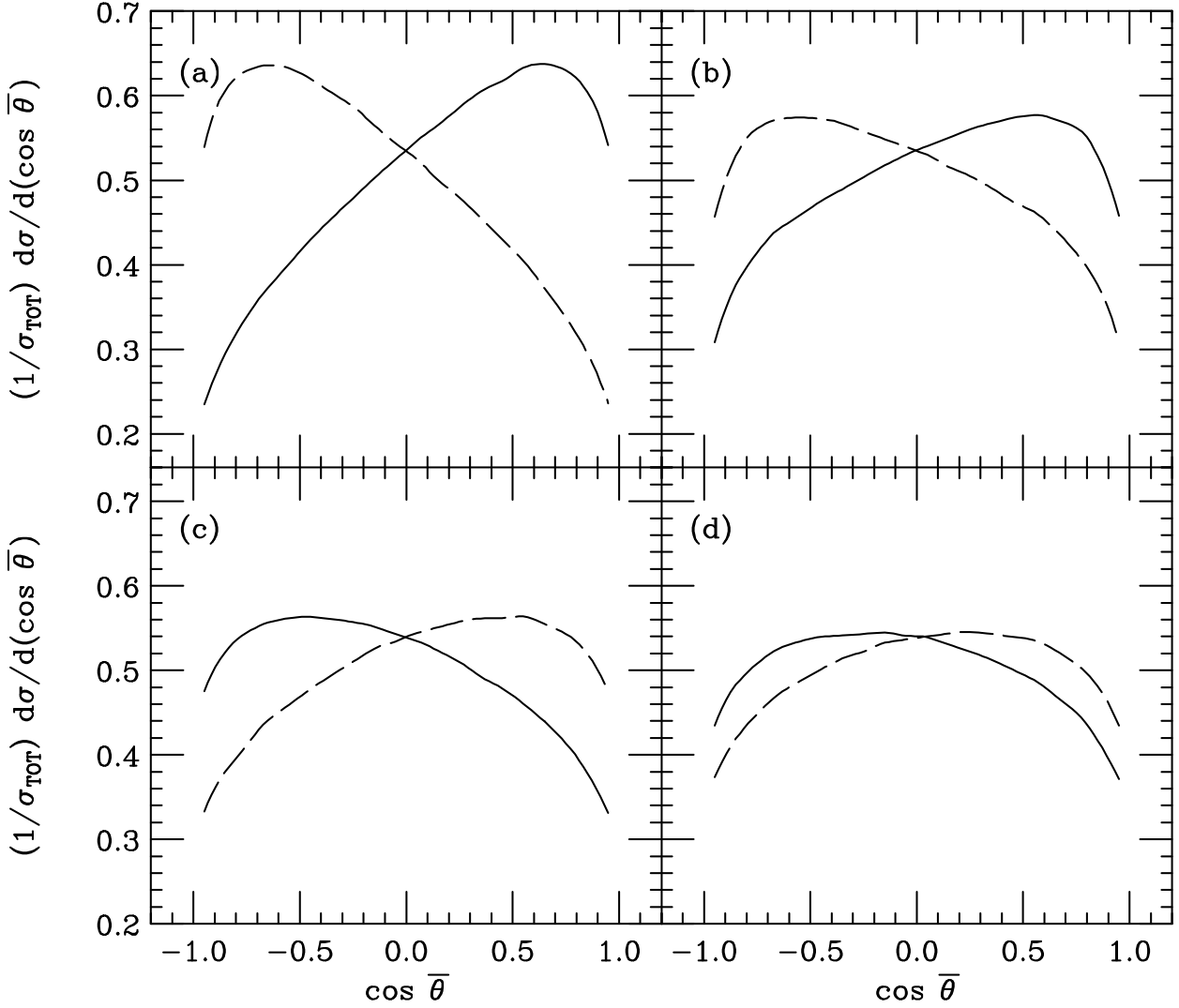}
\vspace{1.0cm}

\caption[]{
Angular correlations in Tevatron $t\bar{t}$ events at 
$\sqrt{s} = 2 \TeV$
using the off-diagonal basis.
The data in each plot are divided into spin-``up'' (solid) 
and spin-``down'' (dashed) top quark components, 
determined by using the charged lepton from the $t$ decay in (a)--(c), 
and  the $b$-quark in (d).  
Plotted are the angular distributions with respect to the 
$\bar t$ spin axis in the $\bar t$ rest frame 
for the following $\bar t$ decay products:
(a) the charged lepton, 
(b) the {\it ``d''}-type quark,
(c) and (d) the $\bar{b}$-quark.
}
\label{CorrPlot}
\end{figure}

In Fig.~\ref{CorrPlot} we show results of a first-pass Monte Carlo
study of the correlations at the parton level without any
hadronization or jet energy smearing effects included.  We
required all final state particles to satisfy the cuts
$ p_T>15 {\enspace\rm GeV}$,  $\vert\eta\vert < 2$.
Plotted in this figure are the angular distributions for
various antitop decay products for samples tagged as
containing spin up or spin down top quarks.  In the absence
of correlations, the two curves in each section of the figure
would lie on top of each other.  Even in the presence of the
cuts, the off-diagonal basis still produces correlations more
than twice as large as those in the helicity basis.

One advantage of the off-diagonal basis which is readily apparent
from Fig.~\ref{CorrPlot} is that the cuts affect the spin-up 
and spin-down samples in a symmetric manner:  {\it i.e.}\ the
two curves remain reflections of each other about 
the point $\cos\bar\theta = 0$.  Consequently, there is no
systematic bias introduced between the two data sets when using
this basis.  The same is not true in the helicity basis:  our
cuts are slightly more likely to exclude left-handed helicity
top quarks than right-handed, and the two angular distributions
acquire different shapes~\cite{MP}.

In conclusion, we have shown that the off-diagonal basis
of  Parke and  Shadmi enjoys several advantages over the
more traditional helicity  basis when applied to a study of
angular correlations in top quark pair production at the
FNAL Tevatron.     Firstly, the vanishing of the amplitude
for the production of like spin top pairs from quark-antiquark
annihilation leads to a total cross section consisting of 92\%
unlike spin pairs in the off-diagonal basis.  This is significantly
better than the 70\% unlike helicity pairs obtained in the
helicity basis.   Secondly, most of the non-diagonal terms
in the production density matrix vanish in the off-diagonal basis.
Those  non-diagonal terms which are not zero are suppressed
by a factor of $\beta^2$.   No such simplicity exists in this
matrix in the helicity basis.  As a result, the contributions
attributed to interference terms in the off-diagonal basis are
far less important than those in the helicity basis.   Thirdly,
the larger production asymmetry of the off-diagonal basis translates
into  angular correlations among the $t$ and $\bar{t}$ decay
products which are more than double those in the
helicity basis.  And, finally, the kinds of experimental
cuts which are typically imposed on collider data do not
introduce a systematic bias between spin up and spin down
top quarks in the off-diagonal basis.
The same is not true in the helicity basis.
These advantages make the off-diagonal basis the basis of choice
for $t\bar{t}$ correlation studies at the Tevatron.


\acknowledgements

The Fermi National Accelerator 
Laboratory (FNAL) is operated by Universities Research Association,
Inc., under contract DE-AC02-76CHO3000 with the U.S. Department
of Energy. 
High energy physics research at the University of Michigan
is supported in part by the U.S. Department of Energy,
under contract DE-FG02-95ER40899.


\vspace*{1cm}

\vspace*{1cm}

\vspace*{1cm}

\vspace*{1cm}

\end{document}